\documentclass{ws-procs975x65}

\begin{document}

\title{Renormalization of Newton's constant and Particle Physics}

\author{X. Calmet$^*$}

\address{Physics and Astronomy, 
University of Sussex,  \\ Falmer, Brighton, BN1 9QH, UK \\
$^*$E-mail: X.Calmet@sussex.ac.uk}

\begin{abstract}
We report on particle physics applications of the renormalization group equation of Newton's constant.
\end{abstract}


\bodymatter

\section*{}

Einstein's dream of a unification of all forces of nature including gravity is still very far away, however much progress has been done in understanding how to formulate quantum field theories in curved space-time and in treating general relativity as an effective field theory. We are used to thinking of the Planck scale $M$ as a fundamental scale of Nature, indeed as the scale at which quantum gravitational effects become important. However, this parameter does get renormalized when quantum fluctuations are taken into account. In other words, Newton's constant and hence the Planck mass  are scale dependent like any other coupling constant of a quantum field theory. The true scale $\mu_*$ at which quantum gravity effects are large is one at which
\begin{equation}
\label{strong}
M (\mu_*) \sim \mu_*.
\end{equation} 
This condition implies that  quantum fluctuations in spacetime geometry at length scales $\mu_*^{-1}$ will be unsuppressed.

A Wilsonian Planck mass $M( \mu )=/\sqrt{G_N(\mu)}$ can be introduced. The contributions of spin 0, spin 1/2 and spin 1 particles  to the running of $M( \mu )$  can easily be calculated using the heat kernel method. This regularization procedure insures that the symmetries of the theory are preserved by the regulator. One finds \cite{Vassilevich:1994cz,Larsen:1995ax,Kabat:1995eq,Calmet:2008tn}
 \begin{eqnarray}
 \label{Nrunning}
M ( \mu )^2=M(0)^2 - \frac{\mu^2}{12 \pi} N
\end{eqnarray}
where $M(0)$ is the Planck mass measured in long distance (astrophysical) experiments and where $N=N_0+N_{1/2}-4 N_1$. The parameters $N_0$,  $N_{1/2}$ and  $N_1$ are respectively  the number of scalar fields,  Weyl fermions and  gauge bosons in the theory. Note that this calculation relies on quantum field theory in curved spacetime and does not require any assumption about quantum gravity. Furthermore, as noted in ref \cite{Vassilevich:1994cz}, the contribution of the photon is gauge independent. It is possible to calculate the contribution of the graviton to the running of the Planck mass treating general relativity as an effective theory. One finds \cite{BjerrumBohr:2002ks} that the graviton contribution to the running of the Planck mass has the same sign as that of the vector field. In other words,  quantum gravitational interactions make the Planck mass bigger at high energy. It is worth noticing that if the number of matter fields $N$ is large, quantum gravitational effects are a $1/N$ correction and thus under control since they represent a small correction to the one loop result. This limit has been studied already by Tomboulis \cite{Tomboulis:1977jk} and later by Smolin \cite{Smolin:1981rm}.

\begin{figure}
\begin{minipage}[t]{0.49\linewidth}
\includegraphics[width=5cm]{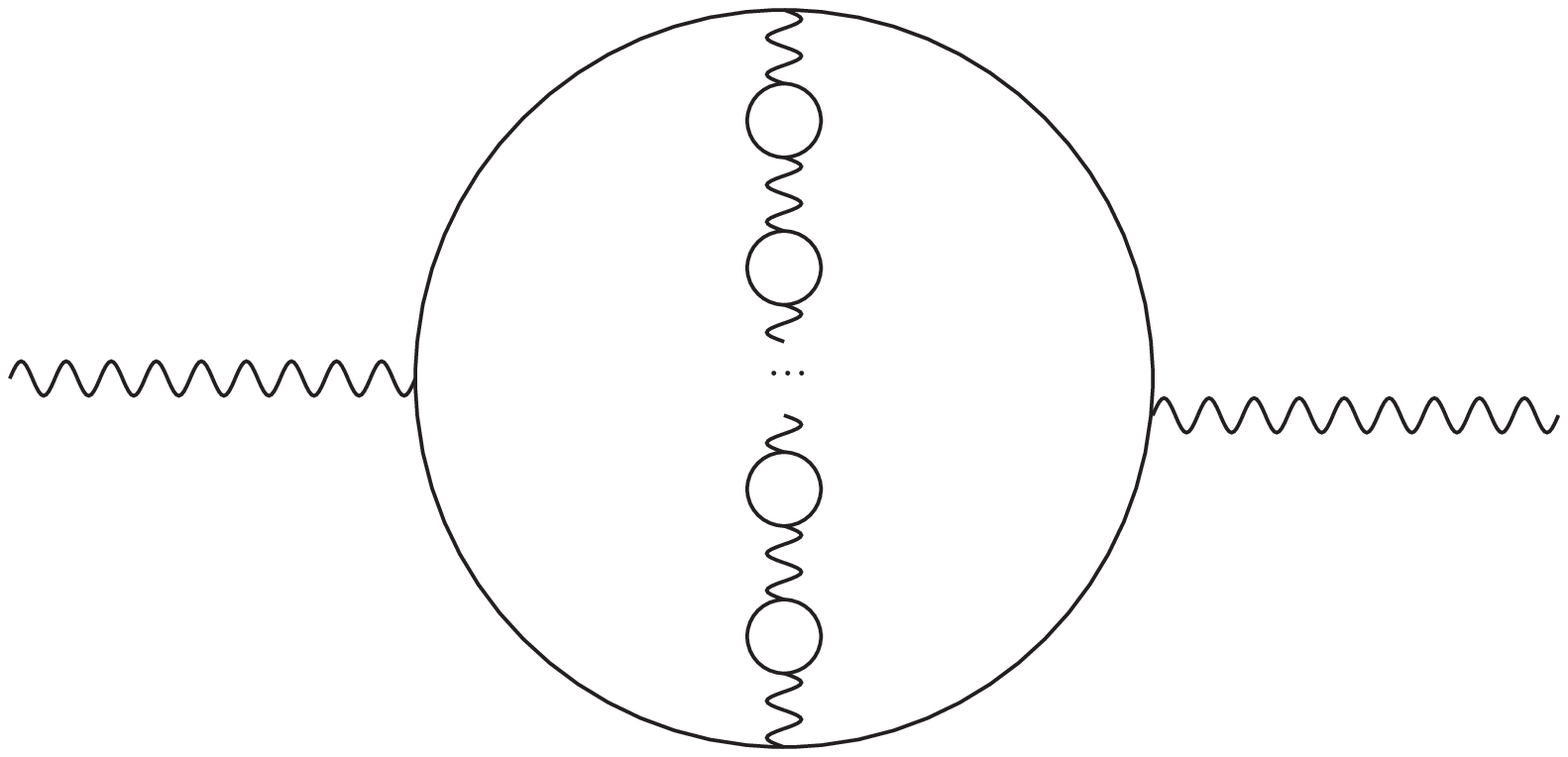}
\caption{\label{FG1}\em
Higher loop contributions to the renormalization of the Planck mass. The wavy lines represent gravitons, whereas continuous lines are matter field loops. For a given number of matter loops, the most important contribution comes from the diagram where a single graviton propagator contains all the matter field loops}
  \end{minipage}
   \begin{minipage}[t]{0.49\linewidth}
\includegraphics[width=5cm]{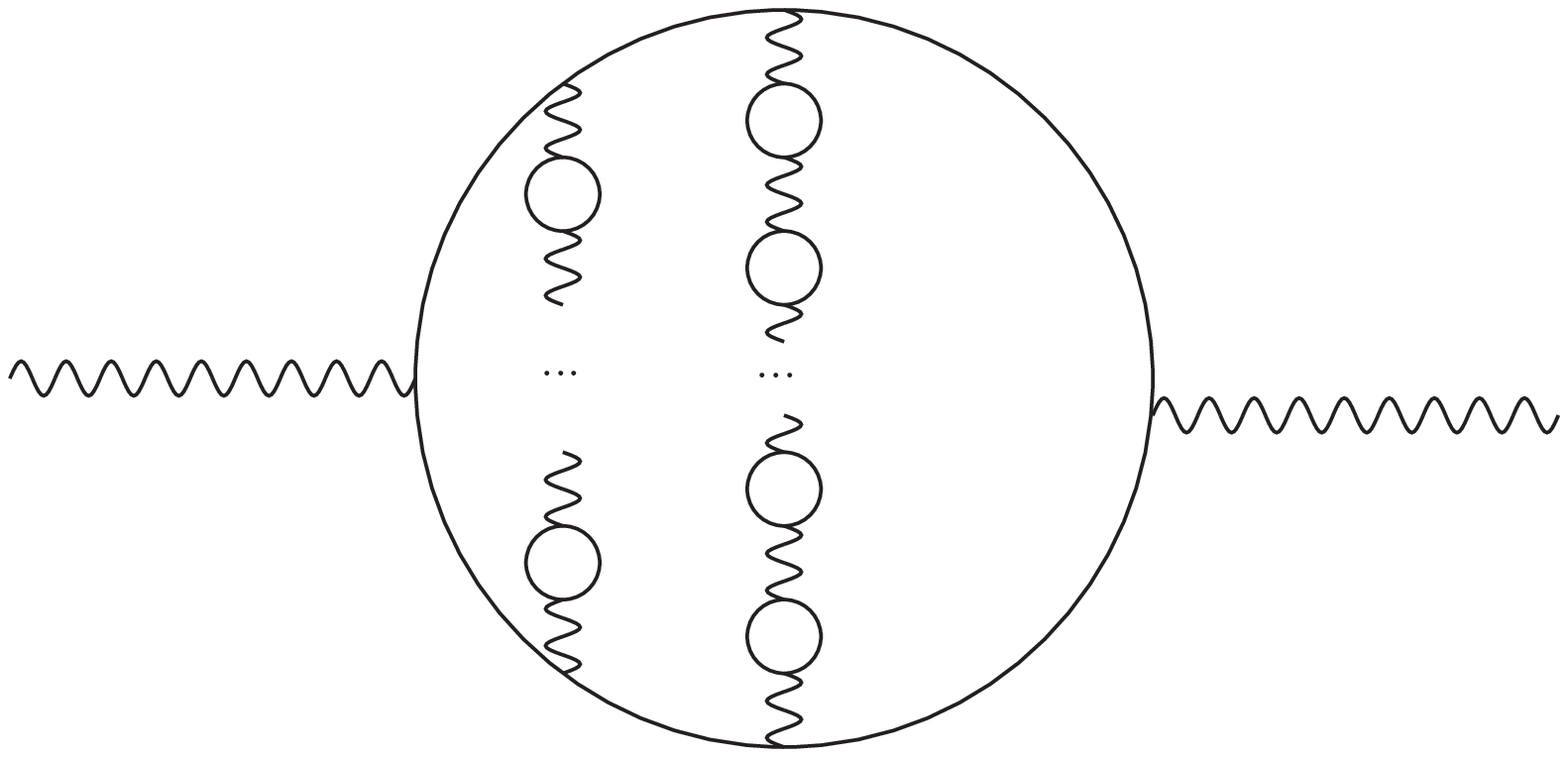}
\caption{\label{FG2}\em
Higher loop contributions to the renormalization of the Planck mass. For a fixed number of loops this topology is suppressed compared to the diagram shown in Figure (\ref{FG1}).}
\end{minipage}
\end{figure}

One may wonder about higher order loop corrections such as diagrams depicted in Figures (\ref{FG1}) and (\ref{FG2}). A back of the envelope calculation shows that diagrams of the type depicted in Figure (\ref{FG1}) lead to a contribution of the type:
\begin{eqnarray}
\sim \frac{1}{(4 \pi^2)^{l+2}}  N^l \left (\frac{\Lambda}{M(0)} \right)^{2l}  \frac{\Lambda^4 N}{M(0)^2}
\end{eqnarray}
where $l$ is the number of matter field loops on the graviton line and $\Lambda$ is a dimensionful cutoff. These contributions are  small compared to the first loop result ($M(0) \sim 10^{18}$ GeV and $\Lambda \sim 10^3$ GeV). In other words if we were able to resum the diagrams on the graviton line, we would obtain a graviton line with a coupling to matter only suppressed $M(\mu_*)$ but not enhanced by $N$.

Diagrams with more graviton propagators (see e.g. Figure (\ref{FG2})), for a given number of matter field loops, are suppressed compared to those shown in Figure (\ref{FG1}):
\begin{eqnarray}
 \sim \frac{1}{(4 \pi^2)^{l+3}}  N^l \left (\frac{\Lambda}{M(0)} \right)^{2l}  \frac{\Lambda^6 N}{M(0)^4}.
\end{eqnarray}

These results can be applied to design models able to solve the hierarchy problem. A large hidden sector of some $10^{33}$ particles of  spin 0 and/or 1/2  reduces the  scale of strong gravity $\mu_\star$ to the TeV region  \cite{Calmet:2008tn}. This could lead to the production of small four dimensional non thermal black holes \cite{Calmet:2008dg} at the LHC or in cosmic ray experiments \cite{Calmet:2008rv} and also to the emission of massless gravitons at the LHC \cite{Calmet:2009gn}.

This running of the Planck mass also has implications for grand unified theories. It has been shown \cite{Calmet:2008df} that in typical supersymmetric  grand unified theories based on for example SO(10), the large number of particles with masses close to the unification scale ($N\sim 1000$) leads to a shift of the strength at which gravity becomes strong. One finds $\mu_\star \sim 10^{17}$ GeV rather than  $\mu_\star \sim 10^{18}$ GeV which implies that operators induced by strong gravitational effects can dramatically impact the unification conditions of the gauge  couplings of the Standard Model. For example SUSY SU(5) may not unify the couplings of the standard model properly. On the other hand, these quantum gravitational effects could also lead to grand unification by the same mechanism in models which naively would not lead to unification of the coupling constants of the standard model such as non supersymmetric SO(10) models\cite{Calmet:2009hp}.

Finally the renormalization of Newton's constant has deep consequences for the validity of linearized general relativity. As shown in ref. \cite{AC}, a comparison  of the scale at which unitarity is violated in gravitational scattering to the scale at which quantum gravitational effects become large leads to a bound on the particle content of a model coupled to linearized general relativity.

\bibliographystyle{ws-procs975x65}
\bibliography{ws-pro-sample}

\end{document}